\documentclass[preprint,3p,onecolumn, number,sort&compress]{elsarticle}
\usepackage{xcolor}
\usepackage{graphicx}
\usepackage{siunitx}
\usepackage{subfigure}
\usepackage[version=4]{mhchem}
\usepackage[
    colorlinks=true,
    linkcolor=blue,
    citecolor=blue,
    urlcolor=blue
]{hyperref}
\usepackage{comment}

\newcommand{\hso}{\ce{H2SO4}}
\journal{Nuclear Instruments Methods A}

\begin{document}

\begin{frontmatter}
\title{\boldmath Hydrofluoric acid-free titanium etching for rare-event searches}

\author[a]{P. Knights\corref{cor1}}
\ead{p.r.knights@bham.ac.uk}
\author[b]{I. Manthos,}
\author[a,b]{K. Nikolopoulos,}
\author[a,c]{G. Rogers,}
\author[a]{D. Spathara\corref{cor1}}
\ead{d.spathara@bham.ac.uk}
\author[a]{P. Walters}
\cortext[cor1]{Corresponding authors}
\affiliation[a]{organization={School of Physics and Astronomy, University of Birmingham},
postcode={B15 2TT},
city={Birmingham},
country={United Kingdom}}
\affiliation[b]{organization={Institute for Experimental Physics, University of Hamburg},
postcode={22761},
city={Hamburg},
country={Germany}}
\affiliation[c]{organization={STFC Boulby Underground Laboratory},
postcode={TS13 4UZ},
city={Boulby},
country={United Kingdom}}

\begin{abstract}
Rare-event search experiments require construction materials with high radiopurity to minimise background contributions. Thanks to its high mechanical strength, low density, machinability, and commercial availability in relatively radiopure forms, titanium is a suitable material for structural elements in rare-event searches. In such applications, a chemical etching stage is typically performed to remove surface contamination or to prepare the surface for further treatment. However, due to its chemical resistance, the etching of titanium conventionally requires hydrofluoric acid, posing serious health and safety concerns that are further exacerbated in deep underground laboratory settings. An alternative approach is proposed, which uses sulphuric acid. Grade 1 titanium samples were etched in 20\% and 40\% sulphuric acid solutions at 20$^\circ$C and 40$^\circ$C for up to 24\,h. The effects of etching were quantified through mass change measurements, surface roughness analysis, and scanning electron microscopy. Sulphuric acid effectively etches titanium, with up to $3.5\,\pm\,0.3\,\si{\milli\gram\per\centi\meter\squared}$ of titanium removed for an unagitated solution of 40\% sulphuric acid at $40^\circ$C for 24\,h. Furthermore, sulphuric acid is shown to be effective at etching at lower concentration and temperature. The formation of a passivation layer during the etching 
may enable control of the total mass removed.
\end{abstract}

\end{frontmatter}

\section{Introduction}
Rare-event search experiments, such as direct dark matter searches and neutrino-less double $\beta$-decay searches, are key to discovering dark matter and understanding the nature of neutrinos. Given that these experiments are searching for extremely rare processes, controlling experimental backgrounds is of paramount importance. As such, unprecedented levels of radiopurity in the respective construction materials are required. This has driven extensive research into the development of radiopure materials, e.g. Refs.~\cite{HOPPE2007486,Hoppe:2014nva, Vitale:2021xrm,  Knights:2025ogz, Spathara:2025hrp, Spathara:2025bfw, NEWS-G:2020fhm}. Where high-strength is required, such as for pressure vessels, titanium is a candidate material. For example, Ti was utilised by the LUX-ZEPLIN experiment~\cite{LZ:2015kxe, LZ:2024zvo}, and is a suitable material for future experiments, e.g. nEXO~\cite{nEXO:2017nam} and XLZD~\cite{XLZD:2024nsu}. 

Although its high tensile strength, machinability, low density, and availability in high purities are important benefits, Ti has some limitations, including limited commercial availability at purities comparable to those of copper. Moreover, during manufacture, some of the tooling material or dust from the environment can be deposited on the surface of the material or implanted in the bulk. As a result, 
radioactive decay products from dust deposits or from gaseous $^{222}$Rn present in the environment can be implanted on the surface during manufacturing.

To mitigate such backgrounds, construction materials can be chemically etched and even subsequently coated with higher-purity layers~\cite{NEWS-G:2020fhm}, preferably after mechanical assembly is complete. The high chemical resistance of Ti makes chemical etching challenging, and conventional methods employ solutions based on hydrofluoric acid (HF) or proprietary mixtures. Proprietary mixtures are not preferable due to uncertainty about their composition, which could represent a potentially unknown source of radioactive contamination. The use of HF is particularly challenging, as it poses serious health and safety concerns. HF is classified as acutely toxic and carries European Regulation (EC) No 1272/2008 hazard statements H300, H310, and H330 (fatal if swallowed, in contact with the skin, or inhaled). As such, extensive controls and specialised training are required for its use. These risks increase further when used in deep underground laboratories. As a result, an alternative etching method, simpler in terms of procedure and chemical mixture, would be transformative to our ability to use Ti in high-radiopurity applications.

Ti is also widely used in medical applications, including dental implants, internal fixation plates, and prosthetics. In Ref.~\cite{BAN20061115}, the surface modification of commercially pure Ti following chemical etching in 48\% \hso~at elevated temperatures is studied.
Aqueous \hso~solutions are already used in deep-underground laboratories for processes such as high-purity copper electroplating, and \hso~is considerably safer and simpler to handle than HF. 
A preliminary study confirmed the feasibility of this approach~\cite{Knights:2025ahk}, motivating the systematic investigation presented here, in which Ti etching in \hso~is studied at concentrations of 20\% and 40\% and temperatures of 20\,$^{\circ}$C and 40\,$^{\circ}$C for times up to 24\,h.

\section{Etching Method}
Sample disk coupons measuring $3.15\,\si{\centi\meter}$ in diameter were cut from a sheet of $0.9\,\si{\milli\meter}$ thick grade~1 Ti using a water jet. The as-received
Ti sheet was formed from a cast billet that was first pressed and then hot and cold rolled to achieve the desired thickness. The coupons were individually treated by immersion in 99.8\% analytical reagent grade ethanol, cleaning in an ultrasonic bath at 30$^{\circ}$C for 30 minutes, and drying at 50$^{\circ}$C for 1\,h. Subsequently, the mass of each coupon was measured.

\begin{figure}[h!]
    \centering
    \includegraphics[width=0.45\linewidth]{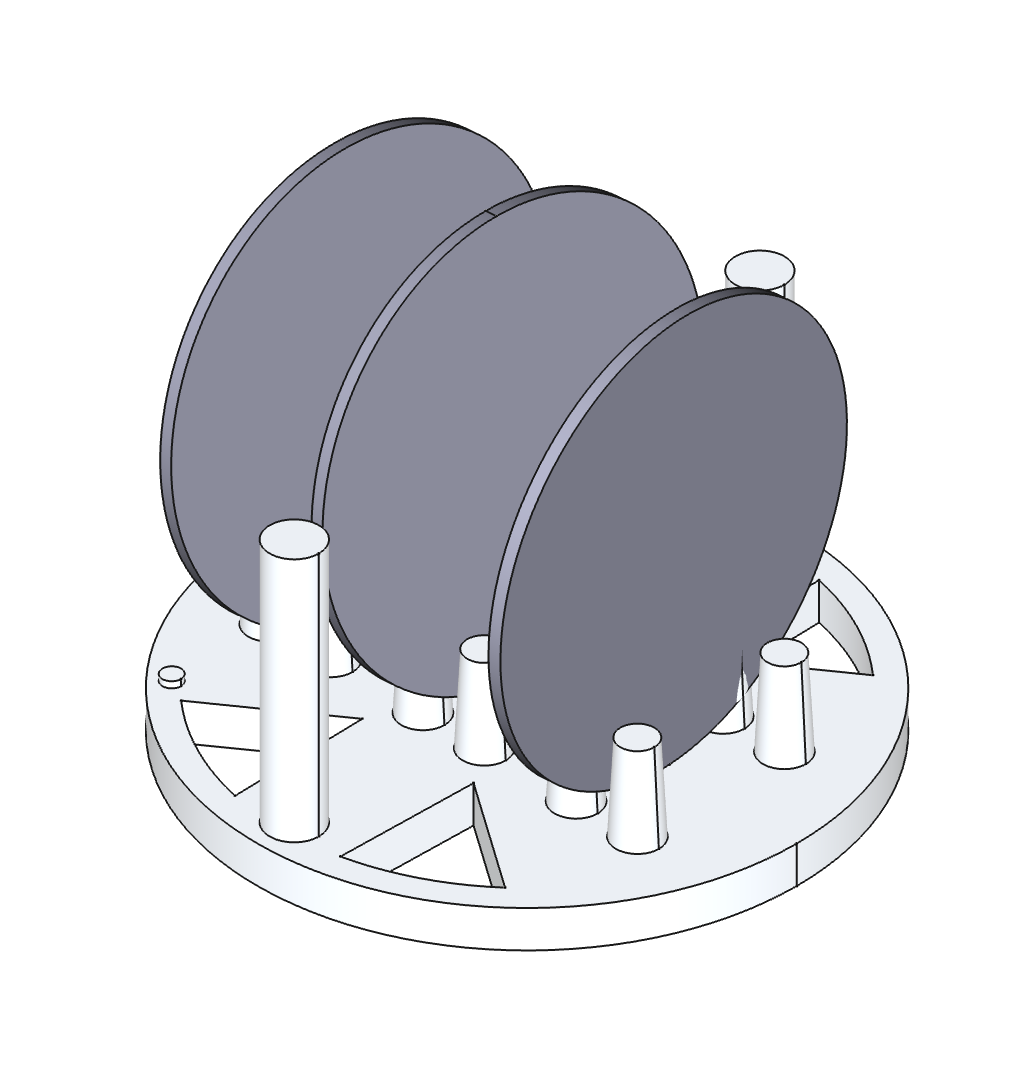}
    \caption{CAD model of a 3D-printed polypropylene fixture, in white, used to hold the coupons upright during etching. The installed Ti-coupons are shown in gray. \label{fig:pp-holder}}
\end{figure}

Coupons were assembled onto a 3D printed polypropylene fixture, shown in Fig.\ref{fig:pp-holder}, designed to keep the coupons upright in the solution and minimise any trapped bubbles formed during etching on the lower surface. The set-up was placed in a $100\,\si{\milli\litre}$ beaker. Solutions of 20\% and 40\% sulphuric acid were prepared using trace metal grade 99.999\% sulphuric acid and type-1 ($18.2\,\si{\mega\ohm}$) water, with $65\,\si{\milli\litre}$ of the solution sufficient to submerge the coupons. To maintain the desired temperature, beakers of solution were submerged in a water bath set at either 40$^{\circ}$C or left at room temperature. The temperature of the water bath was monitored with a K-type thermocouple for the duration of etching. 
The average measured temperatures during etching were $41.3 \pm 1.0\,^\circ$C and $19.3 \pm 0.7\,^\circ$C, respectively. The prepared \hso~solutions were left to preheat in the water bath at the set temperatures for at least 1\,h before submerging the coupons, ensuring that the solution was at the target temperature.

After the target etching time was reached, the coupons were removed and immediately submerged in Type-1 water. The coupons were then rinsed, first with Type-1 water and then ethanol, before drying at 50$^{\circ}$C for 1\,h. The mass of the coupons was then measured again.

\section{Microscopy}
Both optical microscopy and electron microscopy were used to study the effect of etching on Ti surface. For each examined etching scenario, the middle coupon of the 3D printed fixture was selected. Macroscopic studies and roughness measurements were taken using the KEYENCE VHX-7000 digital optical microscope. Because of the unpolished surface both before and after etching, each frame was taken using the 3D mode to achieve adequate focus and capture the surface texture, and therefore a more accurate representation of roughness parameters for each coupon. For the roughness measurements, extended areas of overlapping frames of x1000 magnification were stitched together using the 3D stitching mode.

The Ti surface, before and after etching, was studied using Scanning Electron Microscopy (SEM) in Secondary Electron Imaging (SEI) and Backscattered Electron Imaging (BEI) modes, for topology and composition investigations, respectively. Composition was determined using Energy-Dispersive X-ray (EDX) spectroscopy. These were performed with the JEOL 7000F field emission gun–scanning electron microscope at an acceleration voltage of $20\;\si{\kilo\volt}$ and a working distance of $10\;\si{\milli\meter}$. EDX data were collected with a silicon-drift X-ray detector.

\begin{figure}[h!]
    \centering
    \includegraphics[width=0.45\linewidth]{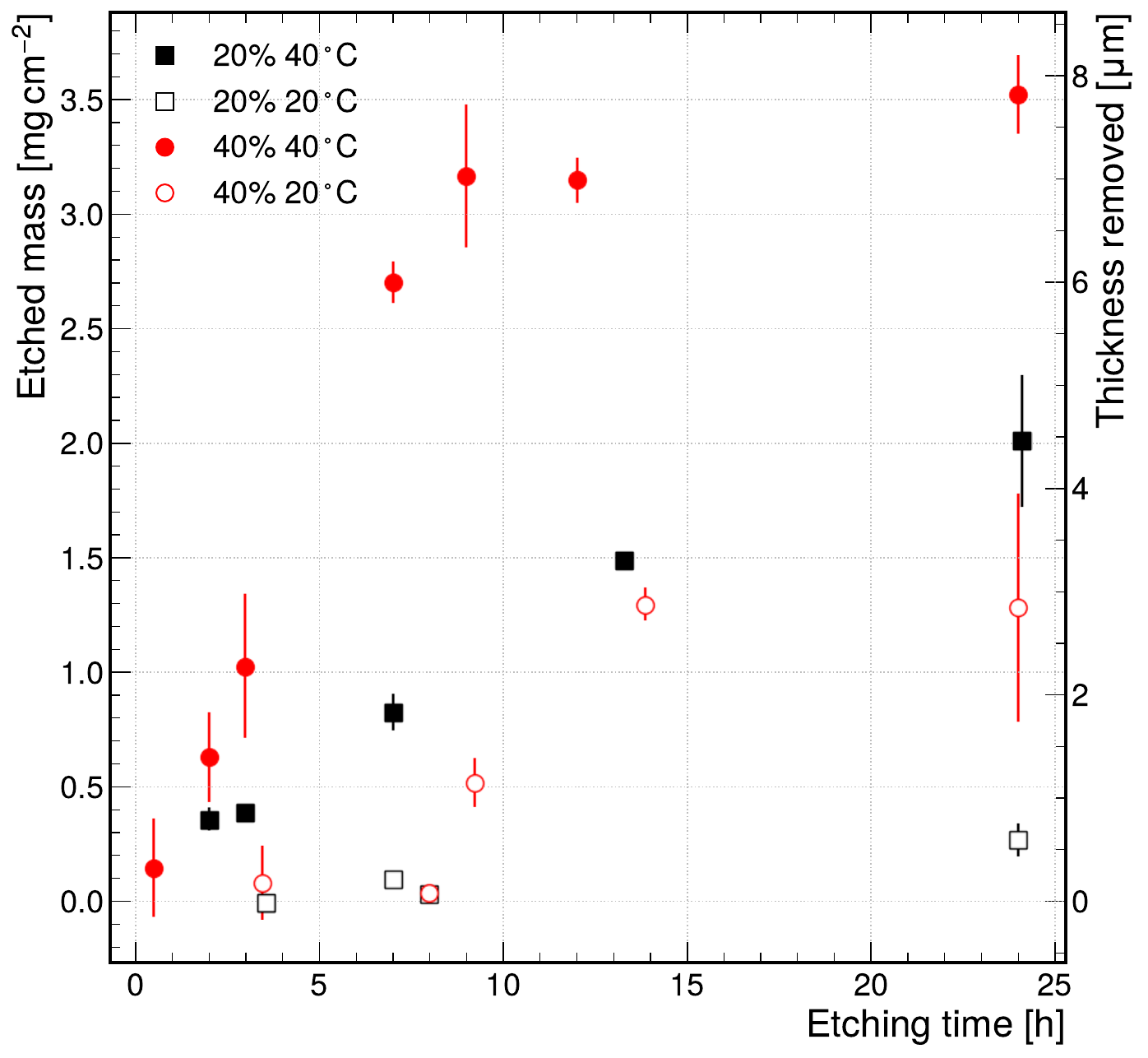}
    \caption{Mass change, and estimated thickness removed, of etched Ti coupons in \hso~solutions of 20\% and 40\% at 40$^\circ$C and 20$^\circ$C. Uncertainties shown are the sample standard deviation of the removed mass of three coupons etched in the same batch at each time and concentration.
    \label{fig:etched-mass-plot}}
\end{figure}

\section{Results and Discussion}
\subsection{Etched mass}
The amount of Ti removed was calculated using the change in the sample mass before and after etching. Assuming uniform removal of Ti and uniform sizing of the coupons, the etched mass per unit area was also calculated using the known diameter and thickness of the coupons. Dividing this by a Ti density of $4.506\,\si{\gram\per\centi\meter\cubed}$ yields the removed thickness. Figure~\ref{fig:etched-mass-plot} shows the variation of the removed Ti mass with etching time, along with the estimated thickness removed.
There is a clear increase in the etching rate with higher temperatures and increased \hso~concentration. 

Initially, the etching proceeds slowly, even in the case of a high temperature of 40\,$^\circ$C and the highest concentration solution examined of 40\%. This is understood to be due to the finite time required to break down the surface TiO$_2$ layer. 
Subsequently, the etching proceeds at a higher rate and finally stops, resulting in a plateau of the removed mass, a behavior which was also observed in Ref.~\cite{BAN20061115}. 

This termination of the reaction is attributed to the formation of a Ti (III) oxide barrier layer coupled with the formation of more complex Ti oxide passivation layers on the coupons surface. Such a layer was clearly visible on the samples, particularly for those most etched as a purple, indicative of Ti (III), film on the coupons. This film is thought to form as a result of the lack of agitation during etching causing the solution in the area immediately surrounding the coupons to become saturated in removed Ti, while  fresh acid was prevented from etching further the coupons surface. With further study this could provide a simple method of controlling the amount of Ti removed during an etching step.

\begin{figure*}[h!]
    \centering
    \subfigure[\label{fig:3a}]{\includegraphics[width=0.45\linewidth]{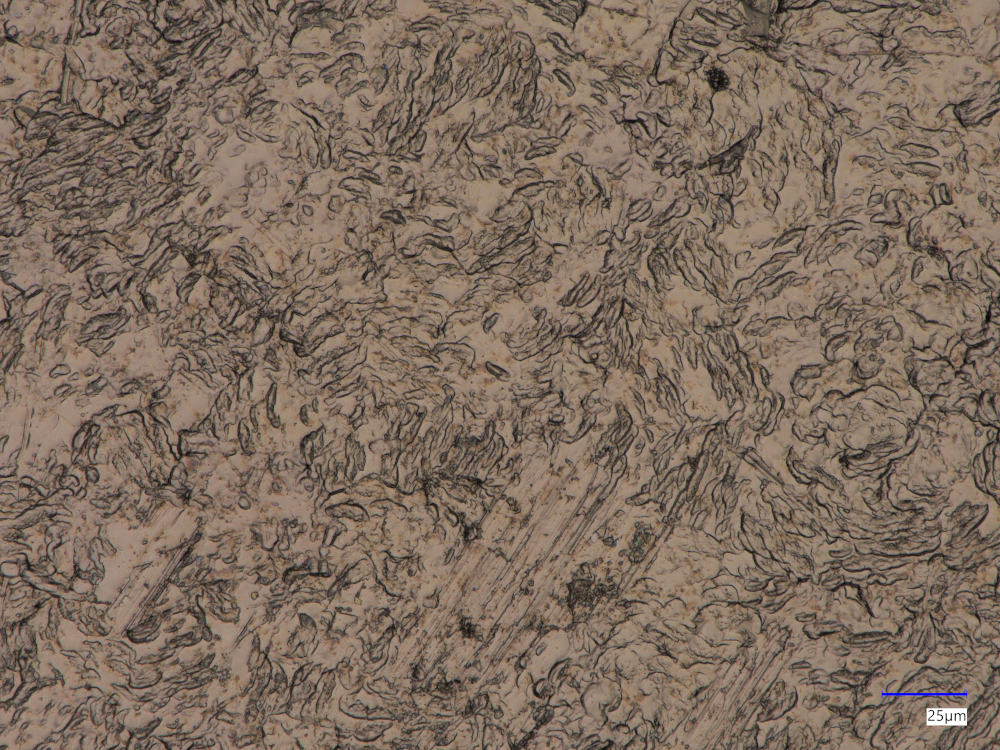}} 
    \subfigure[\label{fig:3b}]{\includegraphics[width=0.45\linewidth]{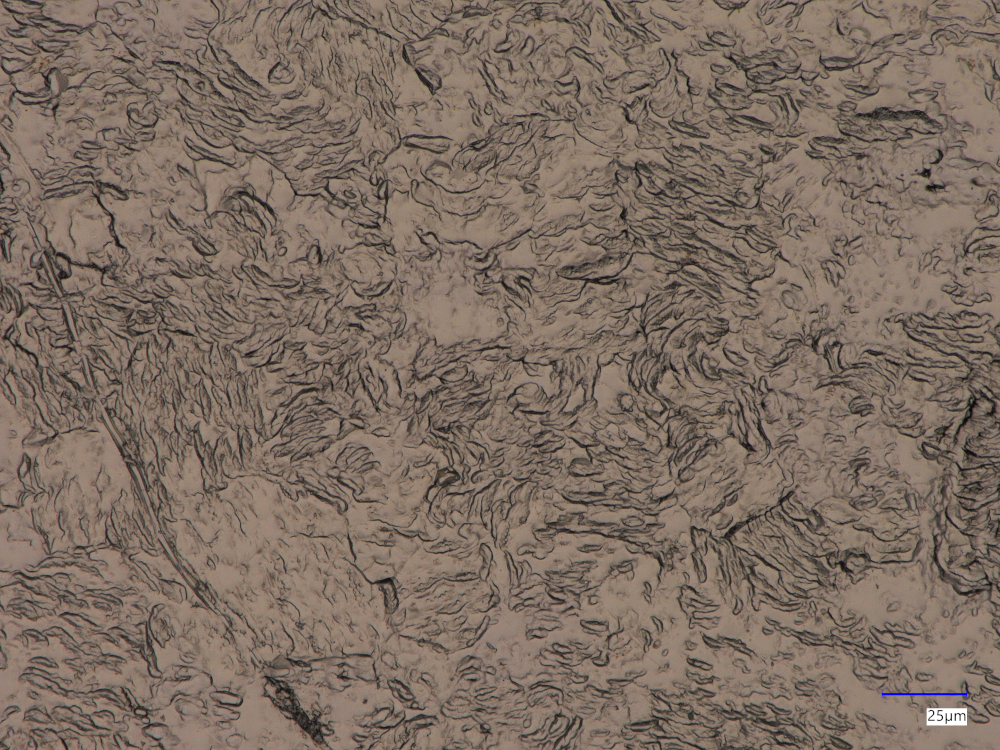}} 
    \subfigure[\label{fig:3c}]{\includegraphics[width=0.45\linewidth]{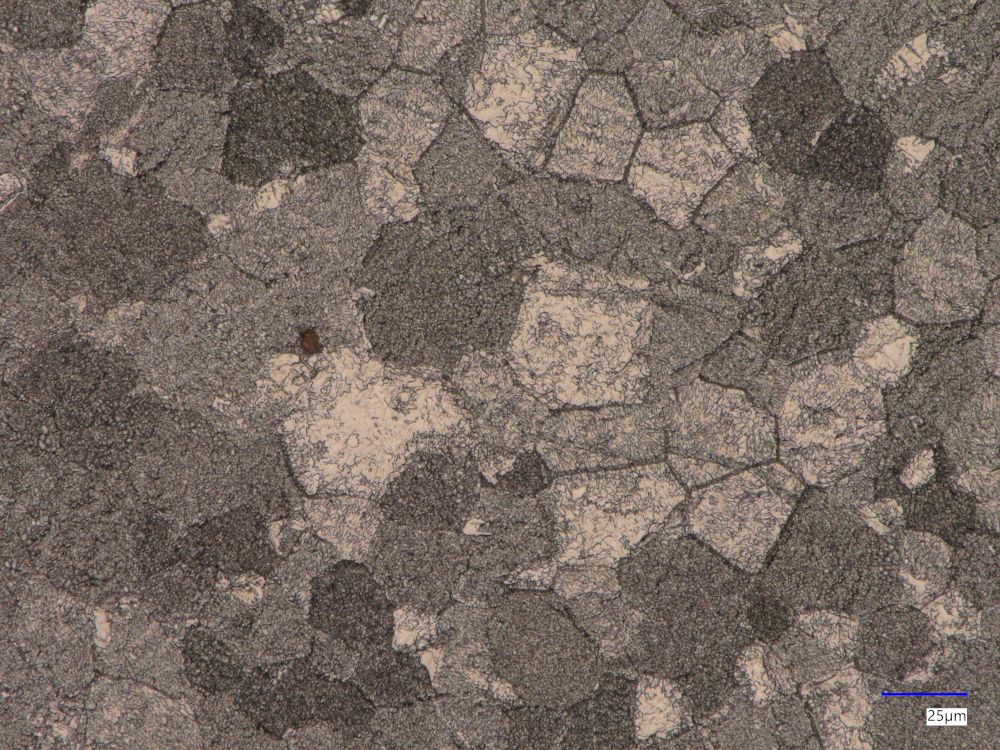}} 
    \subfigure[\label{fig:3d}]{\includegraphics[width=0.45\linewidth]{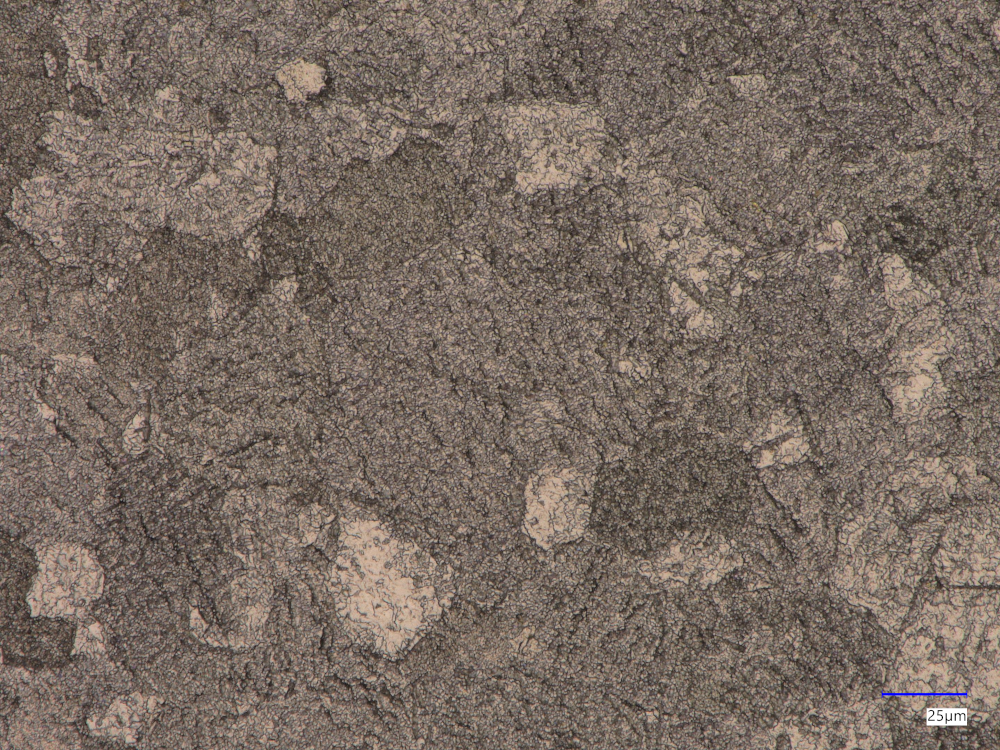}} 
    \subfigure[\label{fig:3e}]{\includegraphics[width=0.45\linewidth]{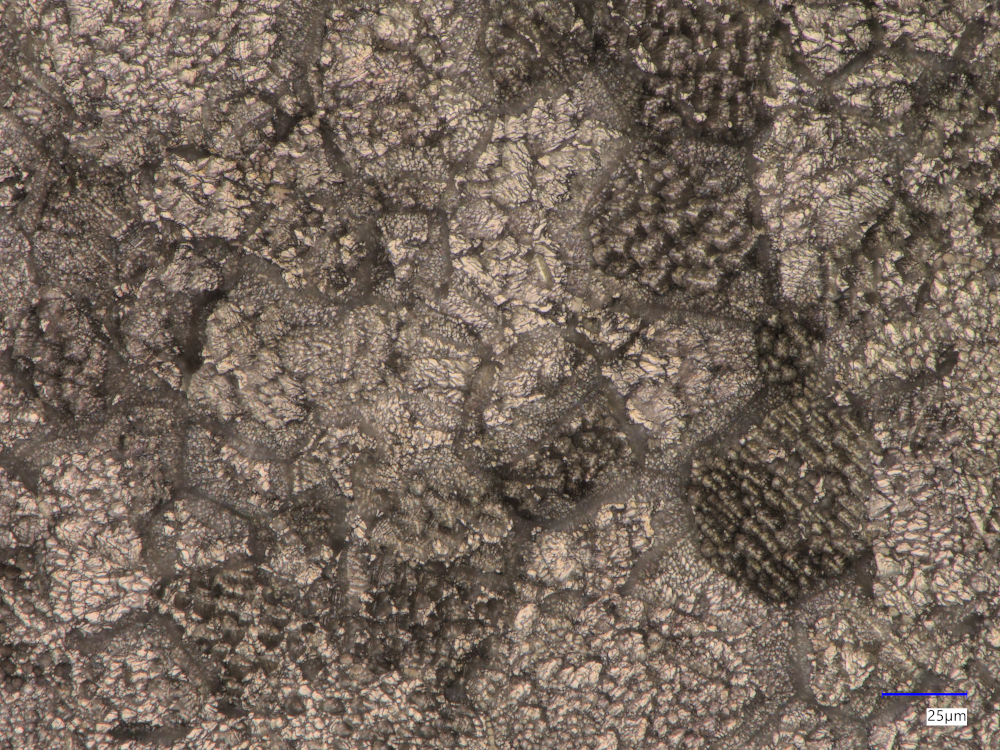}} 
    \subfigure[\label{fig:3f}]{\includegraphics[width=0.45\linewidth]{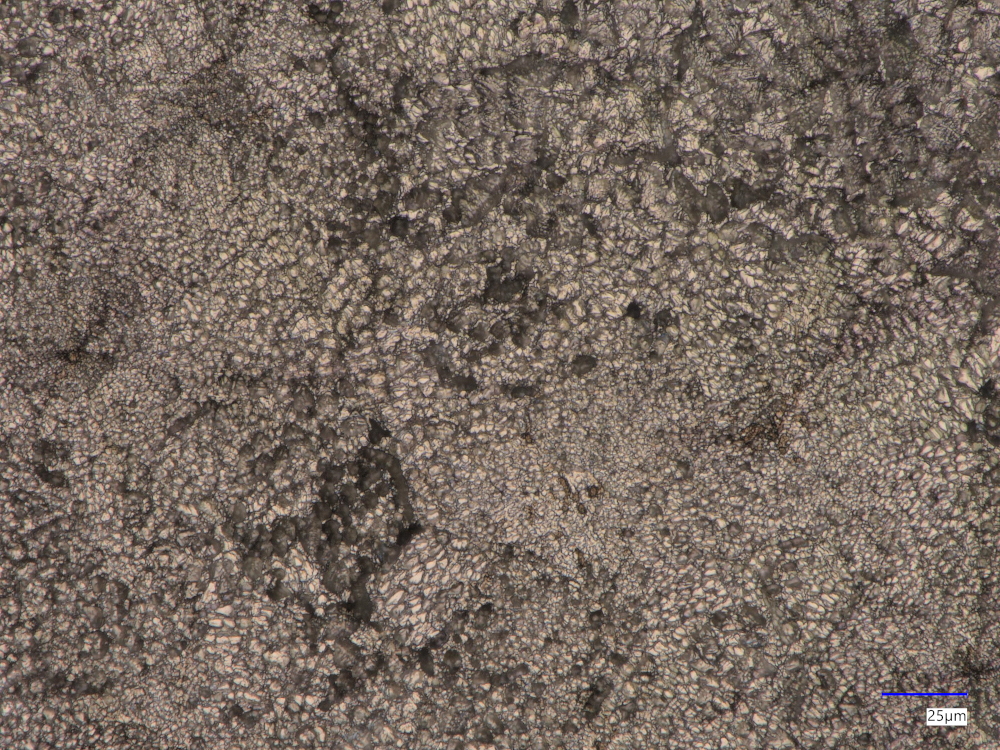}} 
    \caption{Imaging Ti surface x1000 of \subref{fig:3a} a pre-etched surface and etched surfaces, etched using 40\% \hso~at $40\,^{\circ}$C after \subref{fig:3b} 30 minutes, \subref{fig:3c} 2\,h, \subref{fig:3d} 3\,h, \subref{fig:3e} 9\,h and \subref{fig:3f} 24\,h.
    \label{fig:3}}
    \vspace{-0.1cm}
\end{figure*}

\subsection{Surface study}
Figure~\ref{fig:3} shows the Ti surface topology in x1000 magnification before, Fig.~\ref{fig:3a}, and after etching, Fig.~\ref{fig:3}\subref{fig:3b}-\subref{fig:3f}, using 40\% \hso~at $40\,^{\circ}$C, after 30 minutes, 2\,h, 3\,h, 9\,h and 24\,h, respectively. Images were obtained using the VHX-7000 microscope. Since this microscope uses visible light, the lighter areas generally indicate surfaces higher than the darker areas. This was subsequently confirmed using measurements in the 3D mode. It is noted that these observations are specific to the material used for these studies. 

\begin{figure*}[h]
    \centering
    \subfigure[\label{fig:4a}]{\includegraphics[width=0.48\linewidth]{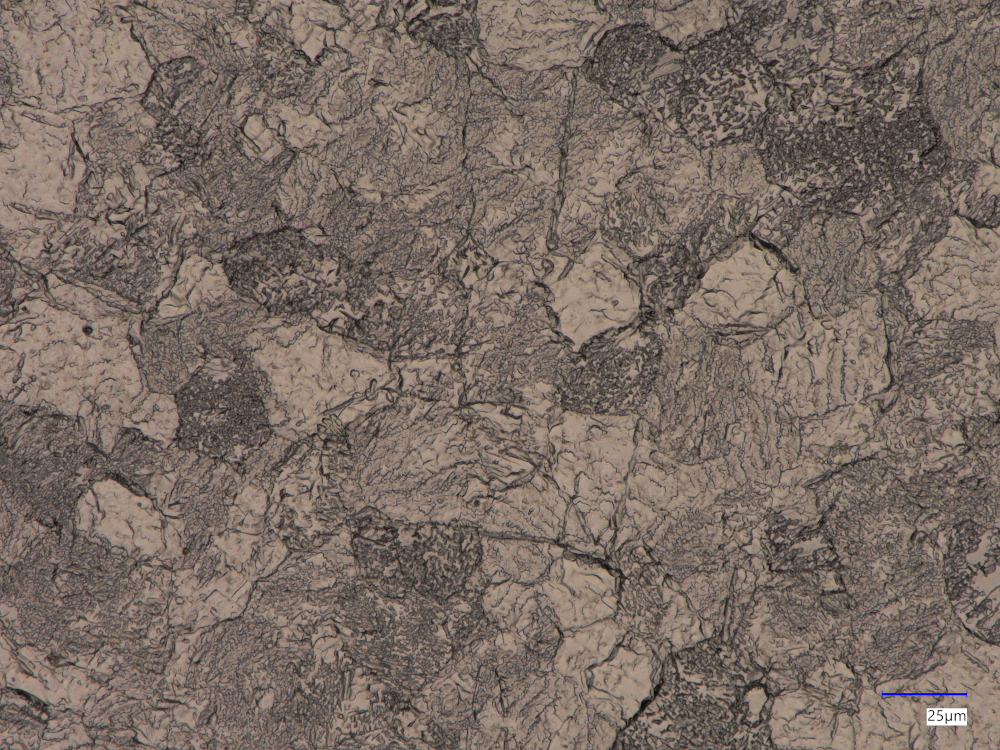}}     \subfigure[\label{fig:4b}]{\includegraphics[width=0.48\linewidth]{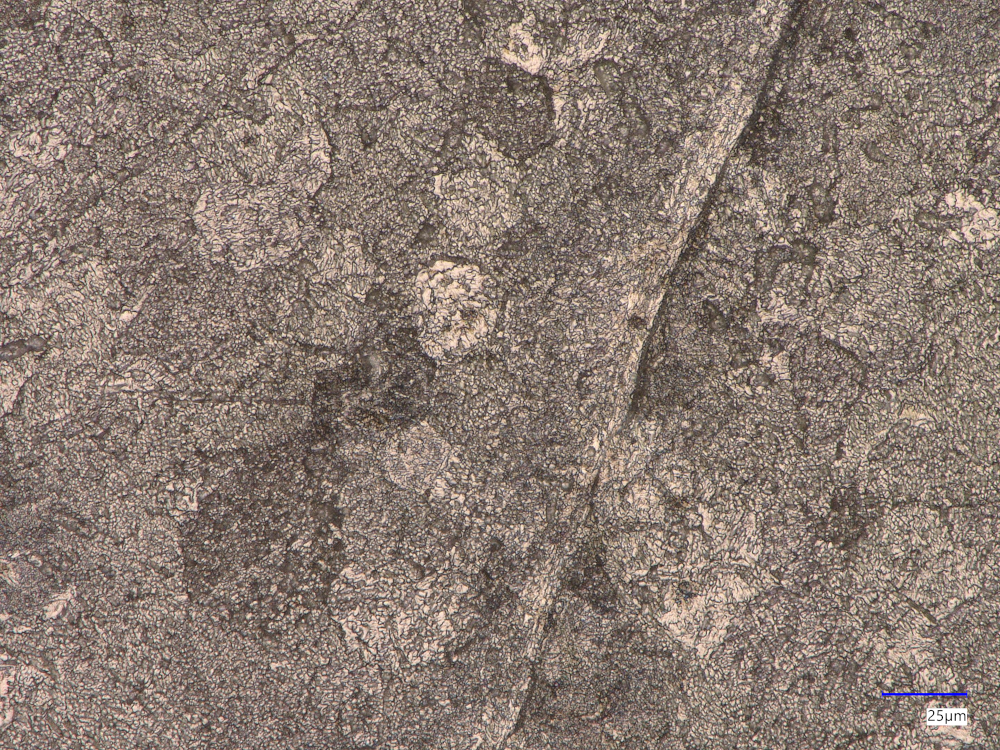}} 
        \caption{Imaging x1000 of Ti coupons, etched at $20\,^{\circ}$C, after 24\,h, using \hso~solution of \subref{fig:4a} 20\%  and \subref{fig:4b} 40\%.
    \label{fig:4}}
    \vspace{-0.3cm}
\end{figure*}

The pre-etched surface reflects the microstructure resulting from hot and cold rolling and heat treatment, exhibiting irregular grains characteristic of the as-processed Ti sheet. After 30 minutes of etching, Fig.~\ref{fig:3b}, the surface appears largely unchanged, though the grain boundaries are somewhat less distinct. At 2\,h, Fig.~\ref{fig:3c}, and 3\,h, Fig.~\ref{fig:3d}, the surfaces look similar to one another. More balanced grain size distribution appears in the 2\,h surface. Additionally, reduction in the clarity of grain boundaries is observed on the 3\,h, 
compared to the 2\,h image. After 9\,h (Fig.~\ref{fig:3e}) grains larger than the 2\,h and 3\,h are seen on the surface, but also much smaller grains appear, sometimes enclosed in these larger grains. The smaller grains are more uniformly dispersed on the 24\,h etched surface.

Generally, as the etching progresses, different configurations of the microstructure are revealed, starting from a dense network of irregular grains (pre-etched and 30-minute) to a more balanced grain size distribution (i.e. 2\,h, 3\,h), to a less balanced grain size distribution; i.e. zones of very small grains are separated by larger grains than previously observed of similar size (i.e. 9\,h and 24\,h). The darker areas in the less distinct boundaries of the 3\,h surface are almost uniform and connected. The lighter grains are smaller in size for 3\,h, compared to 2\,h. This suggests that grains which previously protruded above the surface, as seen in the 2\,h etched sample, have been preferentially removed. Typically, when etching advances on unpolished surfaces, the valleys become deeper. In this case, there is also an indication that the larger grains that sit on top are removed more easily. On the other hand, on the 9\,h etched surface where zones of smaller grains sit on top, there is a more pronounced contrast between the darker grain boundaries and the lighter zones of small grains. 

The topology of the surface for 20\%~\hso~ at 20\,$^\circ$C at 24\,h in Fig.~\ref{fig:4} matches that at 2 or 3\,h for 40\%~\hso~ at 40\,$^\circ$C, in agreement with similarity in the etched mass measurement in each case. 

\begin{figure*}[h!]
    \centering
    \label{fig:6a}{\includegraphics[width=0.9\linewidth]{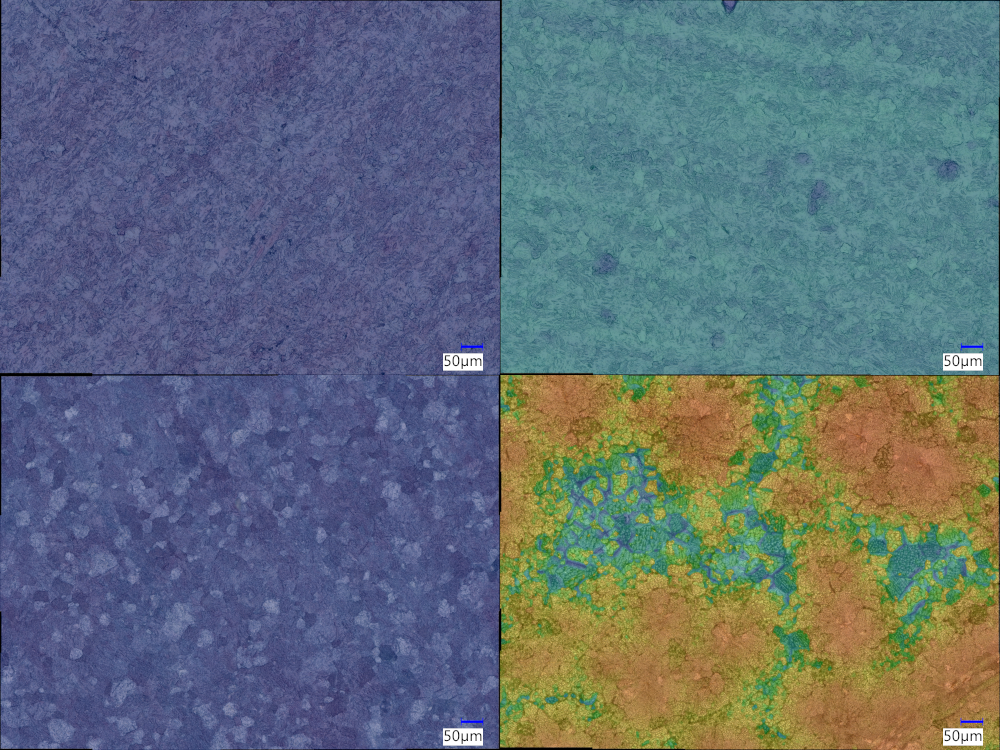}} 
    \label{fig:6-scale} {\includegraphics[width=0.08\linewidth]{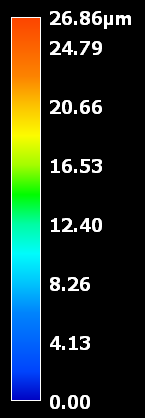}}
    \caption{Imaging x1000, 5x5 frames in 3D mode of Ti pre-etched surface (top left), and etched using 40\% \hso~at $40\,^{\circ}$C after 30 minutes (top right), 3\,h (bottom left), 9\,h (bottom right). The colour scale on the right depicts the height variation of the peaks and valleys.
    \label{fig:6}}
    \vspace{-0.3cm}
\end{figure*}

\begin{figure}[h]
    \centering
    \includegraphics[width=0.45\linewidth]{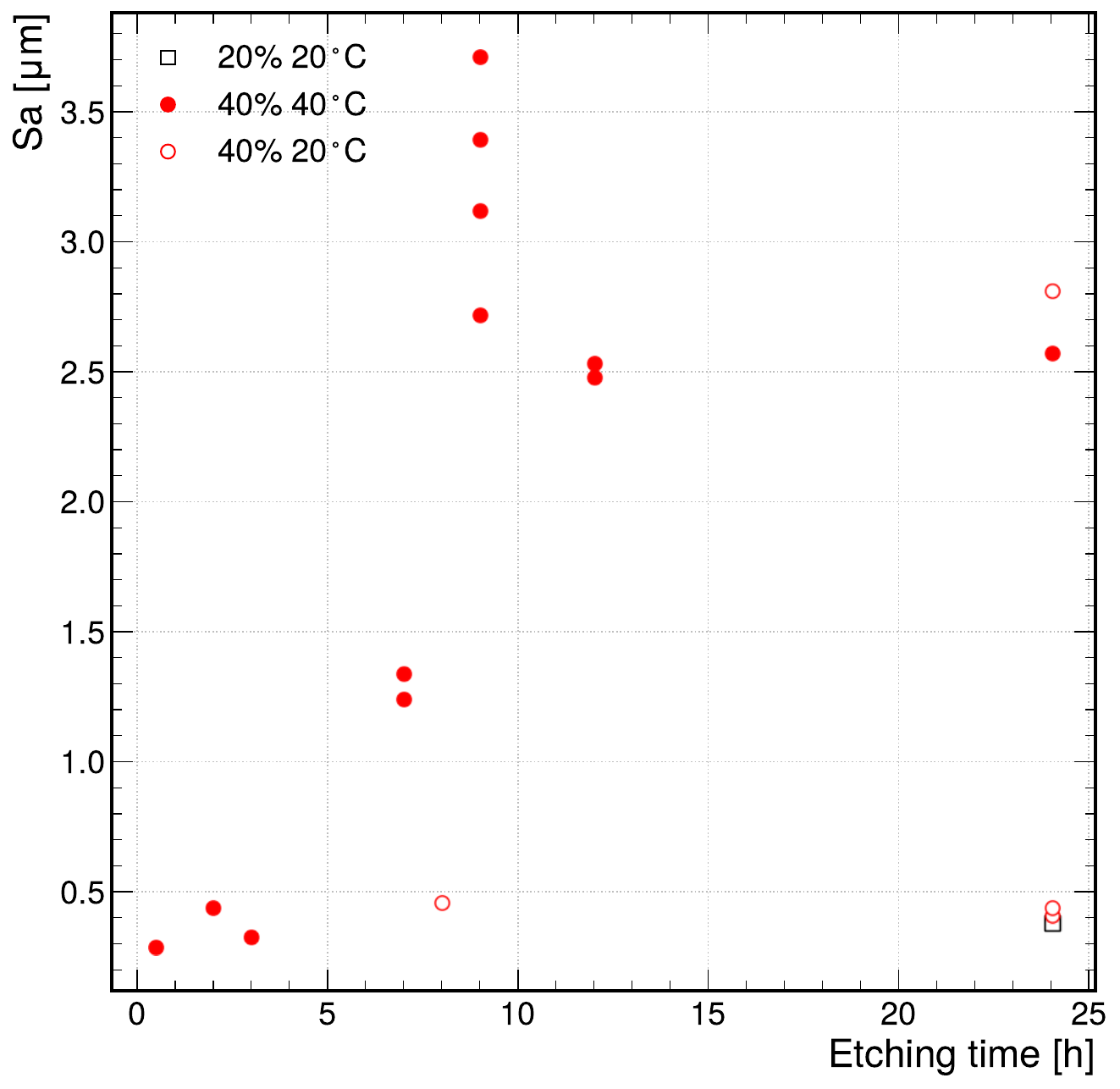}
    \caption{Surface roughness parameter Sa, extracted from KEYENCE VHX-7000 microscope images of Ti coupons etched in \hso 
    solutions of 20\% and 40\% at 40$^\circ$C and 20$^\circ$C and for times up to 24\,h. 
    Multiple measurements at a given etching time correspond to $Sa$ for images at different sites of interest on a coupon.}
    \label{fig:sa}
\end{figure}

Figure~\ref{fig:6} shows extended areas constructed by stitching overlapping frames at x1000 magnification using the 3D mode. This approach enables each frame to be focused independently of local surface roughness. The images show the macroscopic height distribution of peaks and valleys for samples etched in 40\% \hso~ at $40\,^{\circ}$C for 30 minutes, 3\,h and 9\,h, alongside the pre-etched surface. The pre-etched surface shows a peak-to-valley height variation of approximately $3\,\si{\micro\meter}$, which triples after 30 minutes of etching. However, across the approximately $1\,\si{\milli\meter\squared}$ field of view, only a very small fraction of the 30-minute etched surface reaches depths of $3\,\si{\micro\meter}$ below the mean surface level, with an even smaller fraction reaching approximately $9\,\si{\micro\meter}$. A similar restriction of the deepest features is observed at longer etching times, consistent with the plateau in etched mass discussed above.

Surface roughness parameters were obtained for different etching conditions, using the microscope's roughness module. The variation of the roughness parameter $Sa$, defined as the average height of the coupon surface relative to a reference plane, 
with the etching time is shown in Fig.~\ref{fig:sa}.  The values of $Sa$ follow the same trend as the etched mass increases, increasing with etching time. In the case of 40\% \hso~at $40\,^{\circ}$C, a plateau is observed, similarly to the etched mass measured after 12\,h of etching time. The roughness parameters were obtained from images of multiple sites of interest on the coupon. 

The coupons etched in the 40\% solution at 40\,$^{\circ}$C for 7 and 12\,h exhibited similar values of $Sa$ between sites of interest, respectively. On the other hand, the 9\,h etched coupon showed a significantly larger variation of the $Sa$ roughness parameter. This may reflect a transitional stage in the etching process, given the particular microstructure configuration at different stages of Ti removal, as observed when moving from the surface towards the bulk. As etching progresses, the grain boundaries are preferentially attacked, gradually exposing individual grains. 
\begin{figure*}[h]
    \centering
    \subfigure[\label{fig:7a}]{\includegraphics[width=0.45\linewidth]{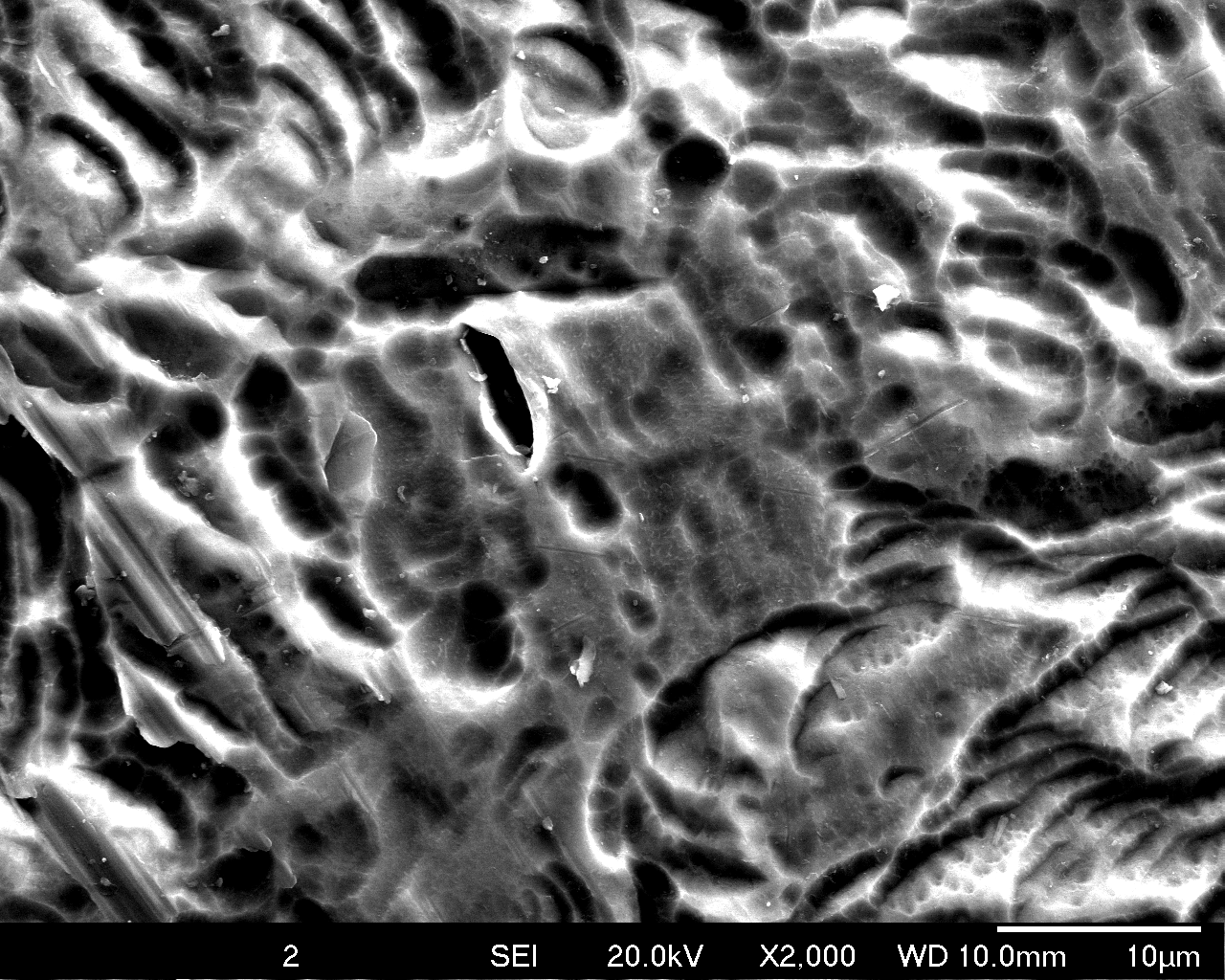}} 
    \subfigure[\label{fig:7b}]{\includegraphics[width=0.45\linewidth]{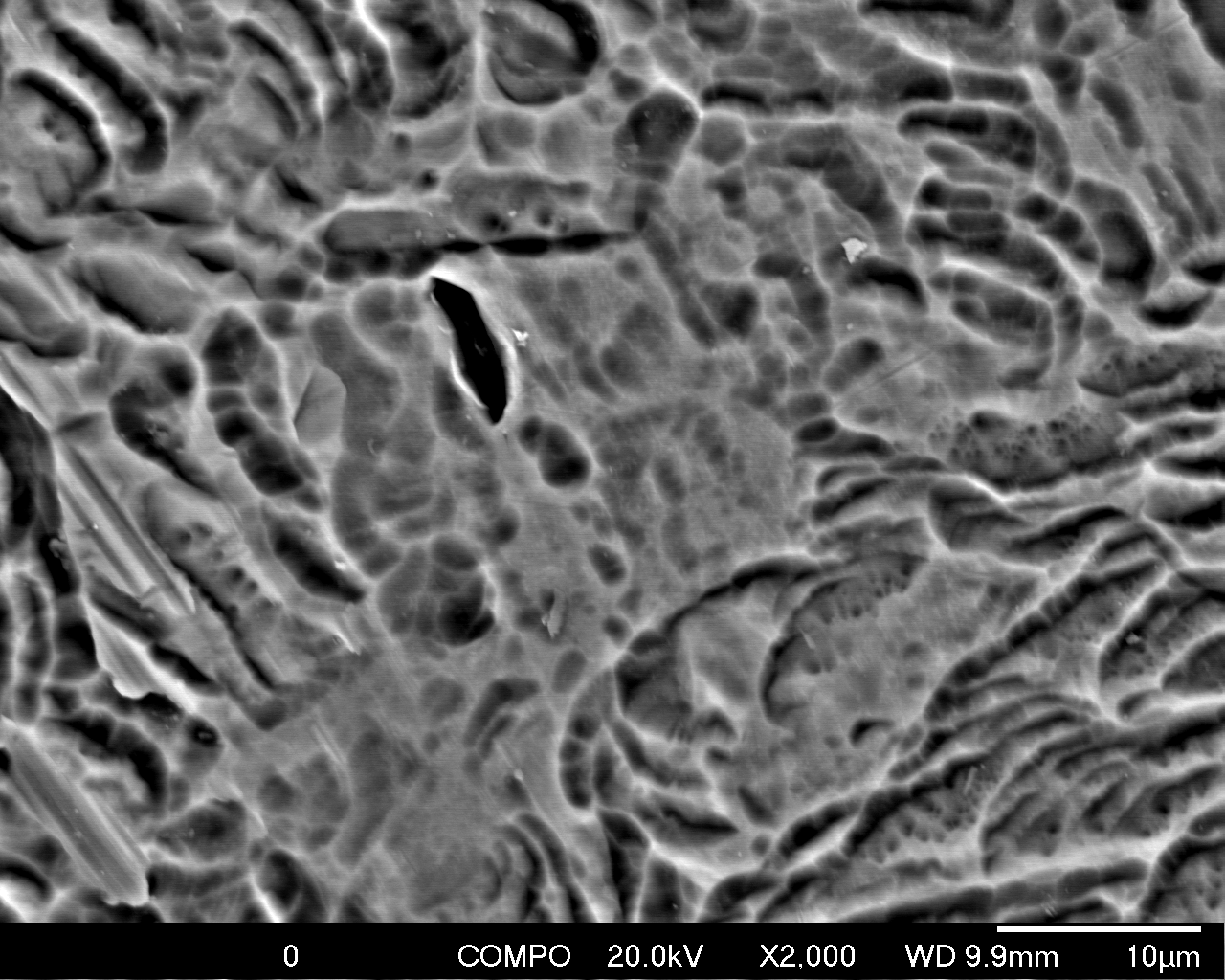}} 
    \subfigure[\label{fig:7c}]{\includegraphics[width=0.45\linewidth]{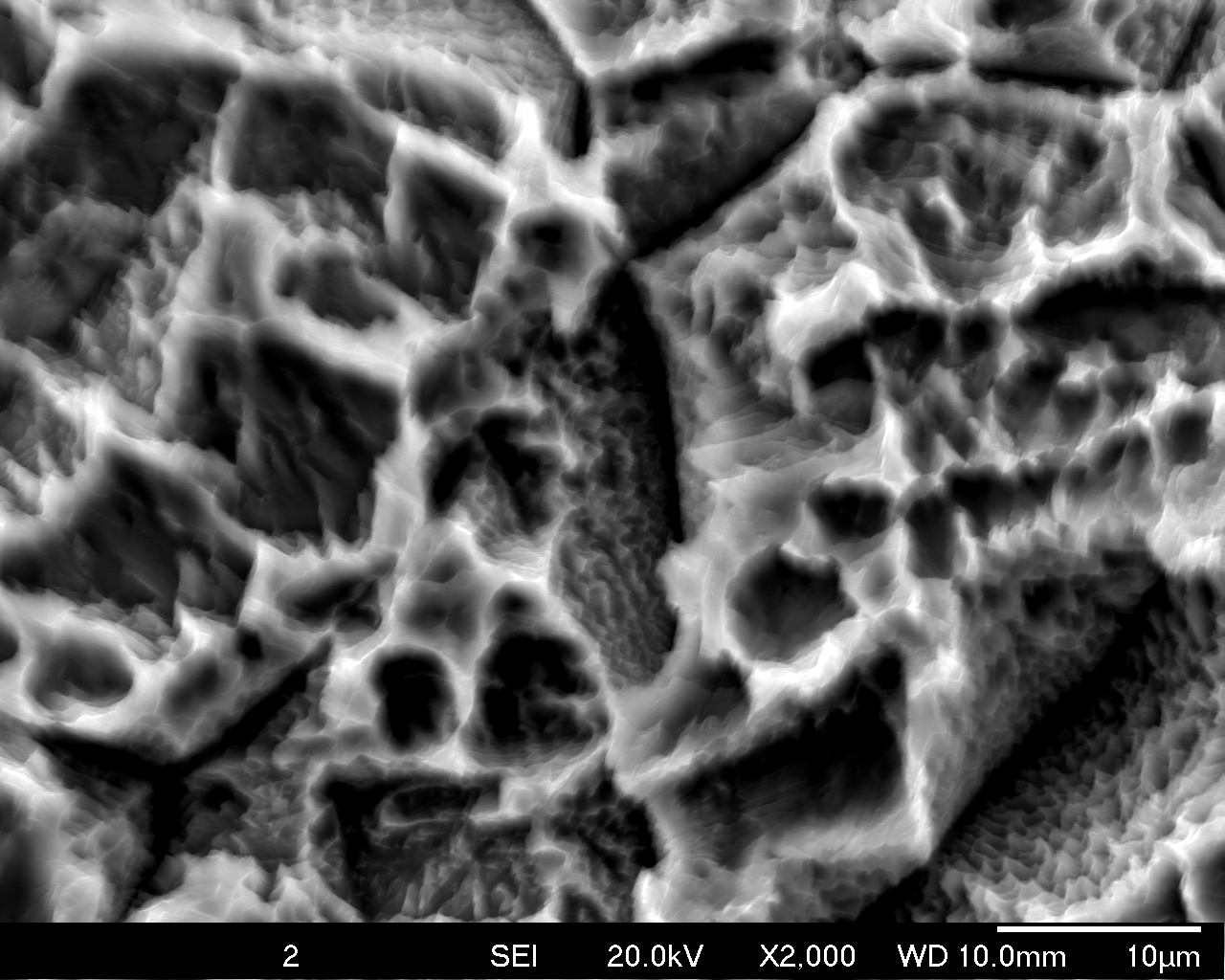}} 
    \subfigure[\label{fig:7d}]{\includegraphics[width=0.45\linewidth]{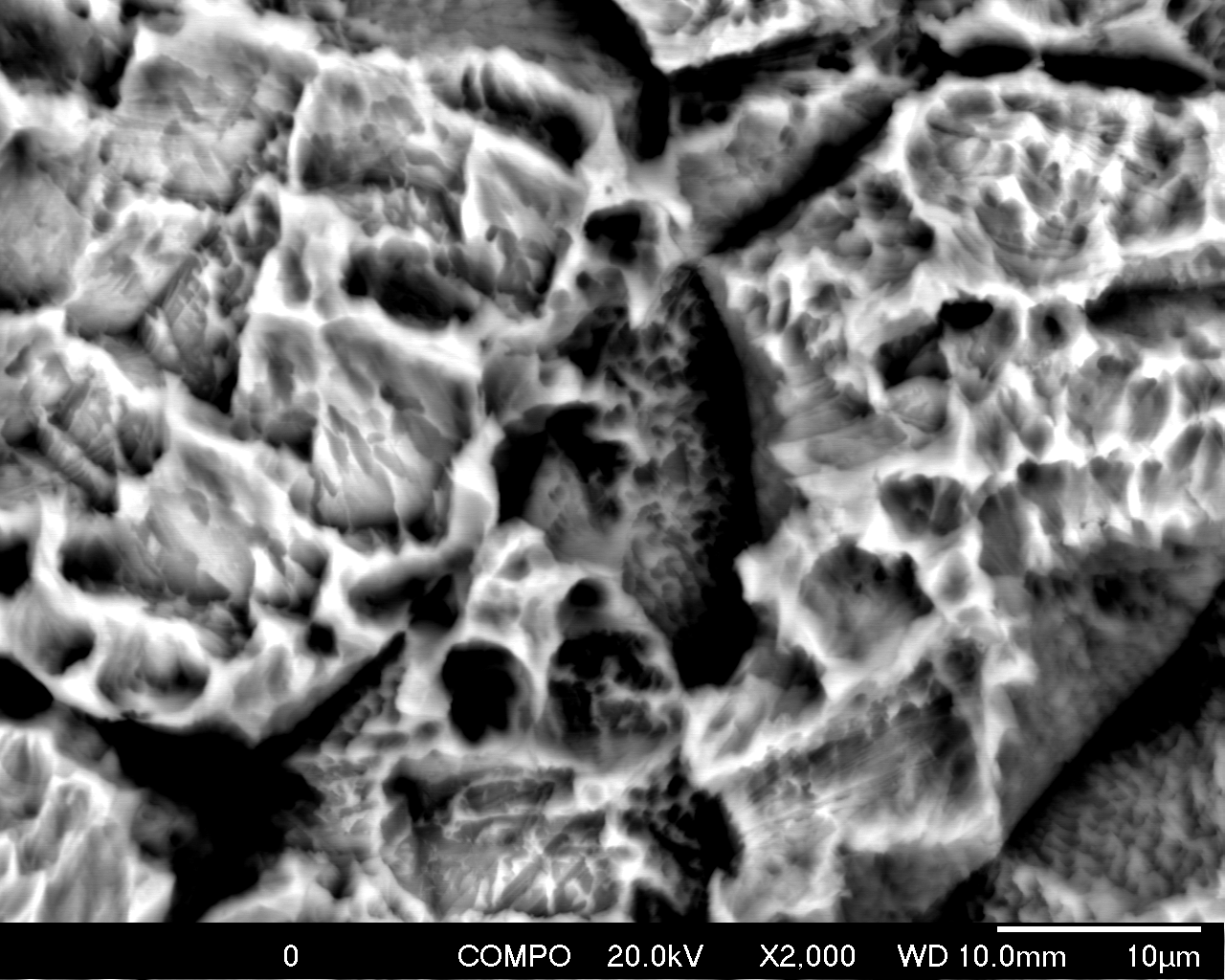}} 
    \caption{Imaging of Ti surface using SEM JEOL 7000F with x2000 magnification: pre-etched  \subref{fig:7a} electron image (SEI), \subref{fig:7b} backscattered image (BSE); and 24h etched using 40\% \hso~at $40\,^{\circ}$C \subref{fig:7c} electron image (SEI), \subref{fig:7d} backscattered image (BSE).
    \label{fig:7}}
    \vspace{-0.3cm}
\end{figure*}
At intermediate etching times, the surface roughness may reach a maximum variation as grains at different stages of exposure coexist on the surface. At longer etching times, the continued removal of material could lead to a more uniform surface. 

\begin{figure*}[h!]
    \centering
    \subfigure[\label{fig:8a}]{\includegraphics[width=0.48\linewidth]{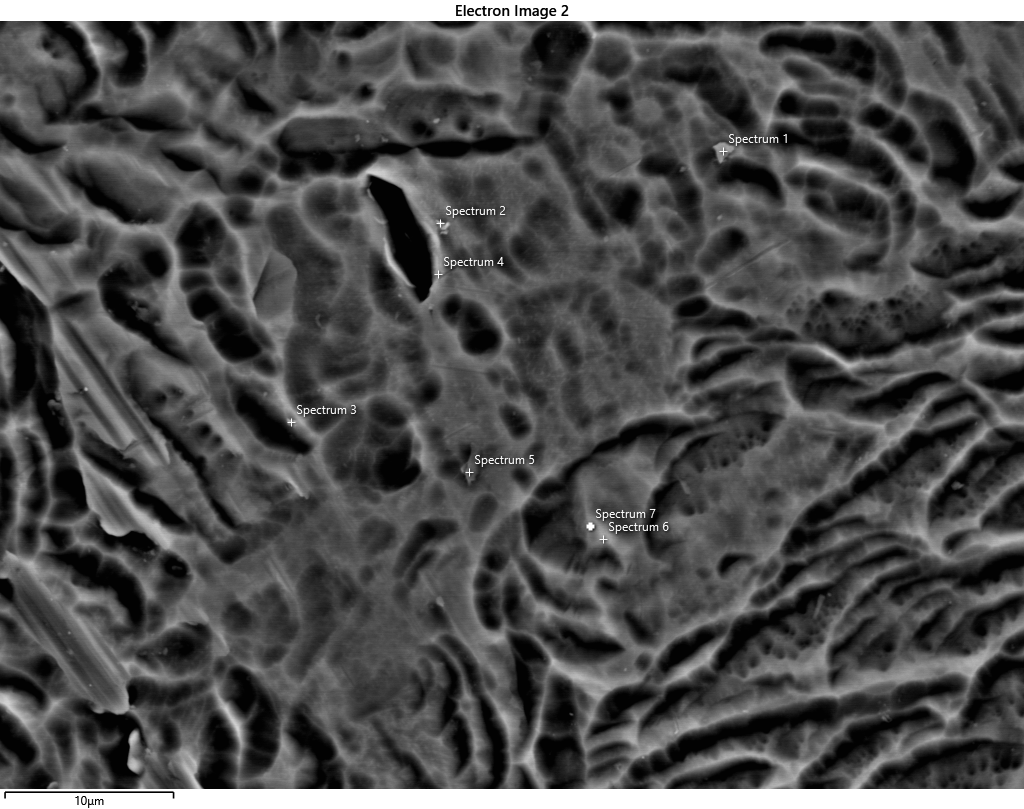}} 
    \subfigure[\label{fig:8b}]{\includegraphics[width=0.48\linewidth]{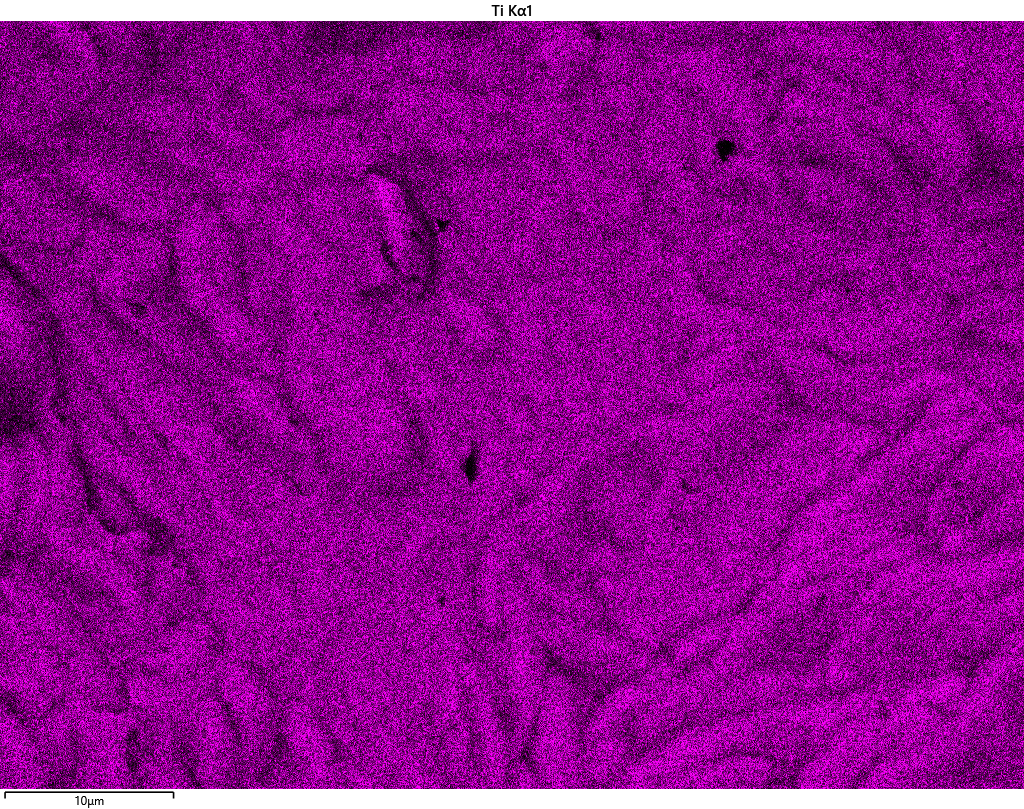}} 
    \subfigure[\label{fig:8c}]{\includegraphics[width=0.48\linewidth]{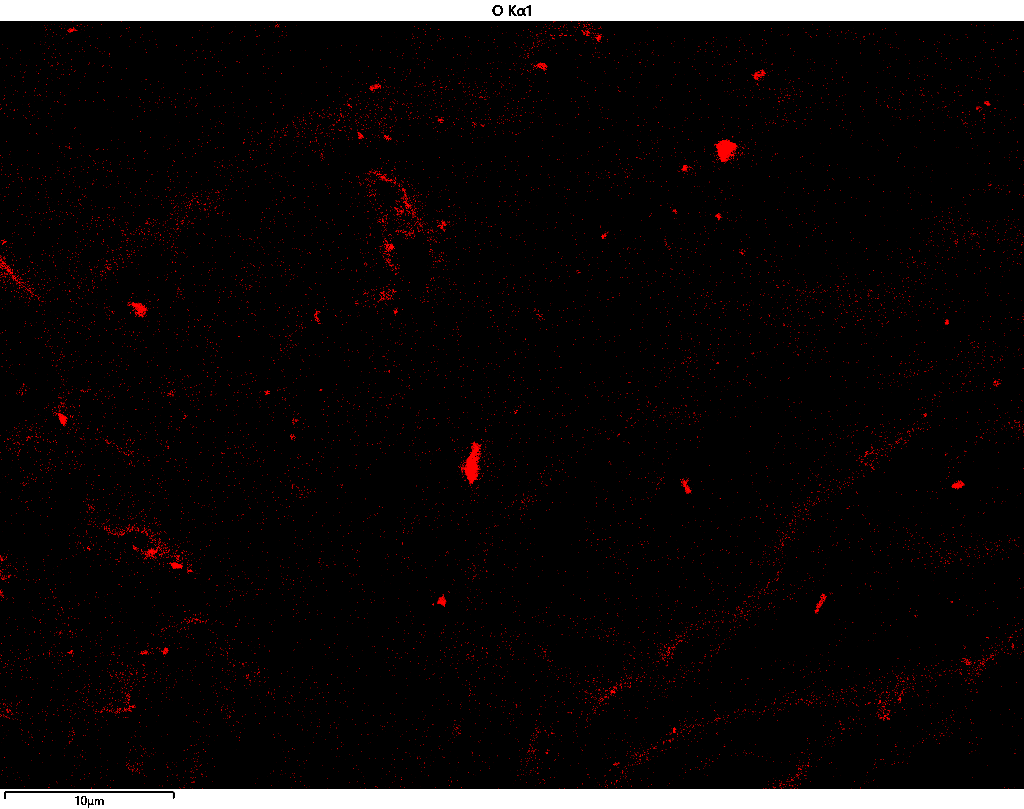}} 
    \subfigure[\label{fig:8d}]{\includegraphics[width=0.48\linewidth]{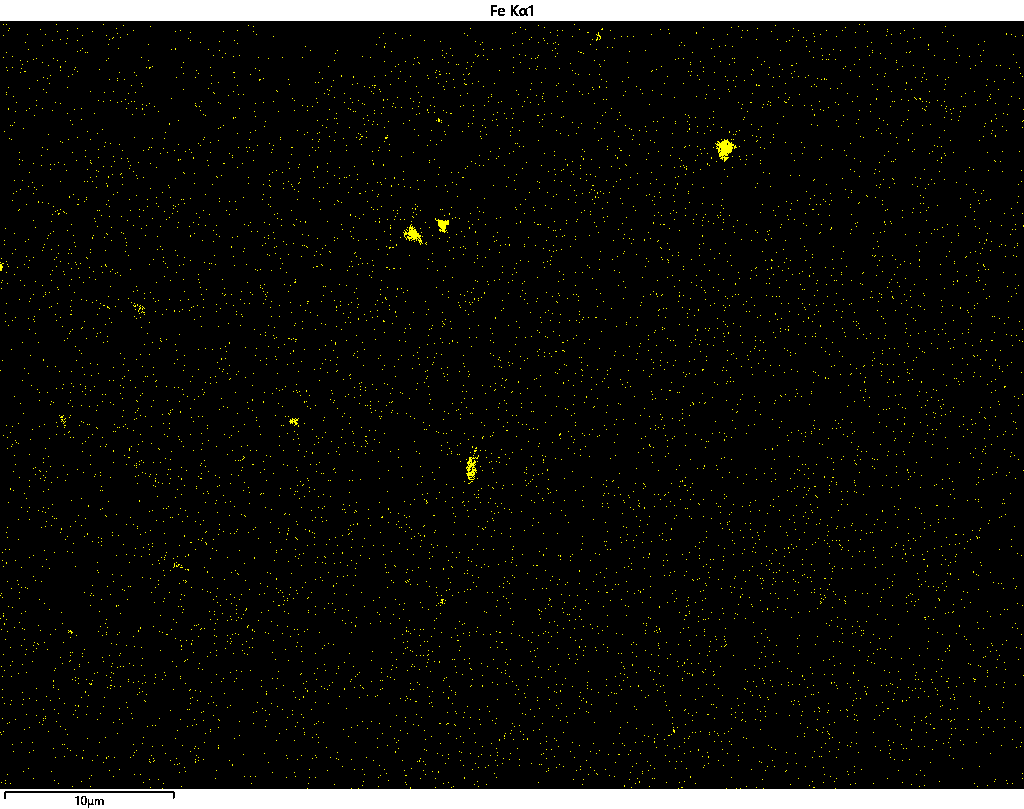}} 
    \caption{\subref{fig:8a} BSE image of Ti surface prior to etching and the EDX maps of \subref{fig:8b} Ti, \subref{fig:8c} O, and \subref{fig:8d} Fe.
    \label{fig:8}}
    \vspace{-0.3cm}
\end{figure*}

Figures~\ref{fig:7a} and \subref{fig:7b} show the results of the SEM imaging of the Ti surface before etching in SEI and BSE modes, respectively. 
The corresponding images of the etched Ti surface after 24\,h, using 40\% \hso~at $40\,^{\circ}$C, are shown in Figures~\ref{fig:7c} and \subref{fig:7d}, for the SEI and BSE modes, respectively. To detect other elements in addition to Ti, EDX maps and EDX point measurements were obtained, as presented in Figure~\ref{fig:8}. 
This revealed localised traces of oxygen and iron contamination on the pre-etched surface.

Surface contamination on the pre-etched surface can be seen by comparing the EDX maps in Fig.~\ref{fig:8} with the BSE image in Fig.~\ref{fig:7b}. The contaminations of iron and oxygen visible in Fig.~\ref{fig:8} are generally also seen in the BSE images as lighter areas. However, as the surface is unpolished, lighter areas in the BSE are also due to the surface topology. EDS maps for the surface post-etching didn't show any element other than Ti.

It is noted that the Ti surface could not be prepared using metallography, i.e. grinding/polishing, to avoid removing information of possible surface contamination. However, as a result, EDX point measurements for contamination cannot be accurately obtained for surfaces that are not adequately levelled due to roughness.
Nevertheless, the X-ray detector used to collect EDX data is capable of capturing high-fidelity images of local contaminations, providing confidence in the observed local contamination in O and Fe from the obtained EDX maps.

\subsection{Discussion of microstructure configuration}
The evolving microstructure observed during etching, illustrated here for 40\% \hso~at $40\,^{\circ}$C 
over 30 minutes to 24\,h, is consistent with the known grain structure resulting from the industrial 
processing of commercially pure Ti sheets~\cite{na2018microstructure, shuai2022effect, 
sun2025microstructure}, and is specific to the material used in these tests. During rolling, 
recrystallisation occurs and large grains are progressively refined into smaller ones, proceeding 
from the bulk towards the external surface. In industrial practice, hot rolling parameters, i.e. pressure, time, and temperature, are carefully controlled to produce a microstructure with 
optimised mechanical properties.

The manufacturing history, of the particular Ti sheets studied, before etching has proven useful to indicate the stage of layer removal. In this case, the very distinct microstructure from the surface towards the bulk can be explained and this helps to identify the stages of etching independently of the solution concentration, temperature and etching time parameters. This becomes important to optimise the etching process and parameters for the specifically manufactured Ti. 

\section{Conclusions}
Titanium etching using sulphuric acid was investigated over a range of concentrations and temperatures for grade 1 titanium sheets. The effects of the etching process on the titanium surface, including surface morphology, roughness, and contamination, were characterized using optical microscopy and scanning electron microscopy. Sulphuric acid was found to effectively etch titanium surfaces even at room temperature. The dependence of etched mass and surface roughness on etching time was quantified. Furthermore, the results indicate that minor surface contamination present prior to etching is removed during the process, an important consideration for low-radioactivity applications.

This hydrofluoric acid-free etching method is particularly attractive because the etchant consists solely of sulphuric acid and water, making the process significantly safer and easier to implement, especially in underground laboratories where etching of large surfaces and complex geometries may be required. In addition, the simple composition of the etchant facilitates radioactivity screening, which is a critical requirement for rare-event search experiments. The ability to etch titanium at relatively low sulphuric acid concentrations and temperatures further broadens the applicability of this technique.

\section*{Acknowledgements}

The support of UKRI-STFC (No. ST/W000652/1, No. ST/Y509036/1, No. ST/Z000777/1) and the UKRI Horizon Europe Guarantee scheme (PureAlloys – EP/X022773/1) is acknowledged. The support of the Deutsche Forschungsgemeinschaft (DFG, German Research Foundation) under Germany’s Excellence Strategy – EXC 2121 “Quantum Universe”-390833306 is acknowledged. The studies using SEM, and EDX  took place at the Electron Microscopy Group Facility, University of Birmingham.

\bibliographystyle{elsarticle-num}
\bibliography{biblio.bib}

\end{document}